\documentstyle[12pt,epsfig]{article}
\newcommand{\beq}{\begin{equation}}
\newcommand{\eeq}{\end{equation}}
\newcommand{\bea}{\begin{eqnarray}}
\newcommand{\eea}{\end{eqnarray}}
\def\lsim{\raise0.3ex\hbox{$\;<$\kern-0.75em\raise-1.1ex\hbox{$\sim\;$}}}
\def\gsim{\raise0.3ex\hbox{$\;>$\kern-0.75em\raise-1.1ex\hbox{$\sim\;$}}}
 
\begin{document}
\begin{center}
{\large{\bf On the Gamma Ray Burst Origin of Extremely Energetic Cosmic Rays}}\\
\medskip
{Nayantara Gupta \footnote{nayan@phy.iitb.ac.in}\\Department of Physics, Indian Institute of Technology Bombay, Powai, Mumbai 400 076, India}
\end{center}
\begin{abstract}
Air shower experiments have detected cosmic ray events of energies
upto 300 EeV. Most likely these cosmic rays have originated from
compact objects. Their exact sources are yet to be identified. It has
been suggested before that gamma ray bursts are possible sources of
ultra-high energy cosmic rays. 
The two models of gamma ray burst emissions most often discussed are the internal and external shock models.
We have calculated the proton spectrum above 60EeV from all gamma ray bursts distributed upto a redshift of 0.02 in the internal shock model assuming redshift and luminosity distributions consistent with observations, log normal distributions for their values of Lorentz factors, variability times and duration of bursts. Within the external shock model we have calculated the proton flux above 60EeV from all nearby gamma ray bursts assuming log normal distributions in their values of total energies, Lorentz factors at the deceleration epoch and compared with the observed data. We find that gamma ray bursts can produce cosmic ray proton flux comparable to the flux observed by the Pierre Auger experiment both within the internal and external shock models. We have also studied the  dependence of the maximum proton energies and the cooling breaks in the proton spectrum on the various parameters like Lorentz factor, energy of the GRB fireball, variability time (in case of internal shocks), ambient particle density (in case of external shocks). Our results are important to understand how the various observable parameters determine which mechanism e.g. $p\gamma$ interactions, synchrotron cooling of protons will dominate over one another inside these sources.     
\end{abstract}

\section{Introduction}
The field of ultra-high energy cosmic rays (UHECR) enriched with the recent
data from the Pierre Auger (PA) experiment \cite{auger1} has emerged
as a very exciting area of research. The highest energy cosmic ray
event so far observed was of energy 300EeV, detected by Fly's Eye
experiment \cite{fly's}. Subsequently, many events of energies more
than tens of EeV were detected by AGASA \cite{agasa} and PA
experiments. The origin of these extremely energetic cosmic ray events
has been investigated since their detection. The current data from PA
experiment shows anisotropy and the signature of GZK cut-off
\cite{gzk1,gzk2}. Powerful cosmic ray accelerators {\it e.g.} gamma
ray bursts (GRBs) and active galactic nuclei (AGN) within 200Mpc from
us are likely sources of the events detected above the GZK cut-off
energy \cite{dermer}. A possible correlation of the AGN within 75Mpc
with the cosmic ray events above 60 EeV has been identified by the PA
collaboration \cite{auger2}. Protons can be accelerated to above 100
EeV within GRBs both in the internal shock \cite{wax1,wax2} and
external shock scenarios \cite{vietri1,vietri2}. 
 It was suggested by \cite{wax1} that if all GRBs emit $5\times10^{50}$erg in cosmic rays and the spectral index ($p$) of the accelerated proton spectrum $(dN/dE \propto E^{-p})$ inside them is $p\leq 2$ then they can explain the observed UHECR spectrum. The association of GRBs with UHECR has been further discussed in \cite{wax2}. The energy production in cosmic rays was assumed to be same as the energy production in photons of energy between 20keV-2MeV which was used as $2\times10^{53}$erg in this calculation and spectral index of proton spectrum $p=2$. The local star formation rate was folded with the maximum distance from which GRBs can contribute to the observed proton spectrum at a particular energy and the proton spectrum from individual GRBs to calculate the diffuse proton spectrum. GRBs were identified as possible origin of ultra high energy cosmic rays. 
Within the external shock model it was shown that GRBs can explain the observed
 UHECR spectrum assuming all GRBs emit an average energy $3.3\times10^{53}$ erg in cosmic rays and proton spectral index $p=2.2$. Although a steeper spectrum gave a better fit for the data below $10^{19}$eV but due to the contributions of Galactic cosmic rays this fit can not be considered as better.     
Ultra-high energy (UHE) heavy nuclei can survive from photo-disintegration in the intense photon fields of GRBs in case of external shocks. It may also be possible in case of internal shocks if the shocks occur at relatively large radii \cite{wang}.
 
A few low luminosity (LL) GRBs have been detected before {\it e.g.}
 GRB 060218/SN 2006aj and GRB 980425/SN 1998bw at redshifts of 0.033 and 
 0.0085 respectively. These GRBs show some common characteristics
 like long duration, low isotropic $\gamma$-ray energy and low luminosity.
Their detections at very low redshifts within a comparatively short period of time imply that they are much more frequent than the canonical HL GRBs
\cite{cobb,pian,soder,liang}.
 The emission of UHE heavy nuclei and protons from high and low luminosity
(HL and LL) GRBs have been studied by Murase et al. \cite{murase}
considering internal and external shocks. HL and LL GRBs can emit UHE
protons and heavy nuclei within the shock scenarios considered in
their paper. Within the internal shock scenario LL GRBs are
unlikely to produce 100 EeV protons but they can produce 100 EeV heavy
nuclei.  The values of the GRB parameters {\it e.g.} luminosity,
Lorentz factor, variability time determine the maximum energy of the
cosmic ray particles produced.

 The Galactic and extragalactic magnetic fields
 distort the trajectories of the charged particles and tracing the
 origin of these particles becomes very complicated. The magnetic
 fields are poorly known. For protons above 60 EeV the deflection in
 regular Galactic magnetic field is expected to be about few degrees
 \cite{auger2}. For Fe nuclei of same energy the deflection would be
 26 times higher. The intergalactic magnetic field has been only
 measured at the centers of rich galaxy clusters. It can produce
 deflections from negligible \cite{dolag} to very large values
 \cite{sigl}. For coherence length of about 1 Mpc the intergalactic
 magnetic field may produce root-mean-square deflection of few degrees
 for protons of energy 60 EeV traversing through 100 Mpc
 \cite{auger2}. As it is more difficult to track the heavy nuclei due
 to their large deflections, the protons of energy more than 60 EeV
 are only considered in this paper.
Modeling of gamma ray bursts has remained as one of the most challenging problems in theoretical astrophysics for the last three decades. Although, lot of progress has been made in the past few years, as of now there is no unique model for GRBs which can explain all aspects and characteristics of GRB emissions in different wavelengths. The leading models for emission from these sources are the internal and the external shock models \cite{zm04}.
We calculate the proton fluxes from all GRBs within the internal shock scenario using distribution functions for their values of luminosities, redshifts, Lorentz factors etc. The luminosity distribution functions of GRBs have been used from
\cite{liang} and for redshift distribution we use star formation rate
from \cite{por}. The Lorentz factors, variability times and burst
durations are represented by log normal distribution functions
\cite{nayan}. Considering the external shock model the time dependent and time averaged proton fluxes have been calculated. Our calculated fluxes have been compared with observed data on UHECR.
\section{UHECR Propagation and Spacial Distribution}
Cosmic rays produced from a point source are expected to reach us much
later compared to the photons due to their deflections by the Galactic
and intergalactic magnetic fields \cite{mir1,mir2}. 
If the cosmic ray protons are randomly scattered in magnetic field $B$ of coherence length $\lambda$ then the typical deflection angle is $\theta\sim (D/\lambda)^{0.5}\lambda/R_L$,
 where $D$ is the distance of the source, $R_L=E_p/eB$ is Larmour radius.  
The time delay of cosmic rays can be expressed as $\tau(E)\sim D \theta ^2/4c$. The dispersion in deflection angle $\triangle \theta$ gives a dispersion in time delay of protons $\triangle \tau(E_p)$. If the dispersion in deflection angle $\triangle \theta << \theta$ then the dispersion in time delay is $\triangle\tau(E) << \tau(E)$. In this case only a few HL GRB events are expected during $\triangle \tau(E_p)$. The extremely energetic cosmic ray events from these GRBs will be spacially localized in a small region which is not consistent with the current data from PA experiment. 
 The proton flux from a GRB in observer's frame is
\begin{eqnarray}
\frac{dN^{ob}_{p}(E^{ob}_{p})}{dE^{ob}_{p}}&=&\frac{dN_p(E_p)}{dE_p}\frac{(1+z)cos^2{\theta_a}}{4\pi D^2}\nonumber\\
&\times& exp(-D/[cos{\theta_a} r_{\pi}(E_p,z)])
\end{eqnarray}
where $E^{ob}_p=E_p/(1+z)$ has been used. We are assuming an average
deflection of $\theta_a$ of the cosmic ray protons above 60 EeV.  The
comoving radial distance of the GRB is
\begin{equation}
D=\int_0^z
\frac{c}{H_0}\frac{dz'}{\sqrt{\Omega_{\Lambda}+\Omega_m(1+z')^3}}
\end{equation}
with $\Omega_{\Lambda}=0.73$, $\Omega_m=0.27$ and the Hubble constant
$H_0=71 km sec^{-1} Mpc^{-1}$.  Due to deflection in magnetic field
the average distance traversed by the protons is $D/cos{\theta_a}$.
The ultra-high energy (UHE) protons lose energy during their
propogation due to photopion, photopair production. The mean free path
of energy loss of an UHE proton with energy $E_p$ is
\cite{dermer,stanev}
\begin{equation}
r_{\pi}(E_{p},z)= (1+z)^{-3}\frac{13.7exp[4/E_{p,20}]}{[1+4/E_{p,20}]}
Mpc
\end{equation}
where $E_{p,20}=E_p(eV)/10^{20}eV$. The proton
flux from a GRB per unit energy and per unit time is
\begin{eqnarray}
\frac{dN^{ob}_p(E_p)}{{dE^{ob}_p}{dt^{ob}_p}}&=&\frac {dN_p(E_p)}{dE_p \triangle \tau(E_p)}\frac{(1+z)cos^2{\theta_a}}{4\pi D^2} \nonumber\\
&\times&exp(-D/[cos{\theta_a} r_{\pi}(E_p,z)])
\end{eqnarray}
All the GRBs happening during $\triangle \tau(E_p)$ 
contribute to the diffuse cosmic ray flux and the factor $\triangle \tau(E_p)$ cancels out in the final expression of the diffuse flux.
\section{The Different Time Scales, Cooling Break and Maximum Energies of Protons}
We define the relevant time scales which can be used to calculate the break energy in the shock accelerated proton spectrum and the maximum energies of protons.The wind expansion time scale or dynamical time scale is defined as $t'_{dyn}=\frac{r}{\Gamma c}$ where $r$ is the radius of expansion and $\Gamma$ is the bulk Lorentz factor. The synchrotron cooling time scale of protons in the comoving frame is $t'_{syn}=\frac{6\pi m_p^2c^3}{\beta_p^{'2}\sigma_{p,T}E_pB^{'2}}$ where the ratio of speed of protons to that of light is $\beta_p^{'}\sim 1$, $B'$ is the magnetic field in the comoving frame and $\sigma_{p,T}$ is the Thomson scattering cross section for protons. The cooling time scale due to $p\gamma$ interactions is $t'_{p\gamma}=\frac{r}{\Gamma c f_{\pi}(E_p)}$. 
From the condition $t'_{dyn}=t'_{syn}$ we get the break energy in the proton spectrum due to synchrotron cooling $E'_{p,c,syn}$. The break energy due to $p\gamma$ interactions $E'_{p,c,p\gamma}$ can be calculated from $t'_{dyn}=t'_{p\gamma}$. The cooling break energy in the shock accelerated proton spectrum is close to the minimum of these two breaks
\begin{equation}
E'_{p,c}=min[E'_{p,c,syn},E'_{p,c,p\gamma}]
\end{equation}
The protons are shock accelerated in magnetic field by Fermi mechanism. Their acceleration time scale is $t'_{acc}=\frac{2\pi \zeta E'_p}{eB'c}$ where $\zeta\sim 1$. If we compare $t'_{acc}$ with $t'_{dyn}$, $t'_{syn}$ and $t'_{p\gamma}$ we get the cut-off energies in the proton spectrum $E'_{p,max,dyn}$, $E'_{p,max,syn}$ and $E'_{p,max,p\gamma}$ respectively. The minimum of these three gives the maximum energy of the shock accelerated protons.
\begin{equation}
E'_{p,max}=min[E'_{p,max,syn},E'_{p,max,p\gamma},E'_{p,max,dyn}]
\label{p_max}
\end{equation} 
\section{Proton Spectrum in the Internal Shock Model}
The internal shock model of prompt emission is widely discussed in the
literature \cite{rees2,pilla,frag,bhat,raz1,pe'er1,pe'er2,gupta1}.
Its most successful aspect is its ability to model the complex temporal
 profiles of light curves during GRB prompt emission. The observed temporal behaviour reflects the temporal behaviour of the central engine.
For long bursts the central engine is located deep inside a collapsing star, whose envelop may act as an additional agent to regulate the variability of the relativistic flow. Relativistically moving shells with Lorentz factors $\Gamma\sim 100-1000$ emerge from the core. The faster moving shell catches up the slower one. When the shells collide with each other their kinetic energy is transformed
into internal energy of the fireball. Electrons and protons are shock
accelerated in irregular magnetic fields when the shells collide. The
total energy of the fireball is distributed among the electrons,
protons and the magnetic field where $\epsilon_e$, $\epsilon_p$ and
$\epsilon_B$ denote the equipartition parameters respectively.  The
relativistic electron/proton spectra can be represented by broken
power laws
\begin{equation}
\frac{dN_x(E'_x)}{dE'_x}\propto\left\{ \begin{array}{l@{\quad \quad}l}
{{E'_x}^{-p}} & E'_{x,m}<E'_x<E'_{x,c}\\ {E'_x}^{-p-1} & E'_{x,c}<E'_x
\end{array}\right.
\end{equation} 
in the case of slow cooling where $x=e,p$ for electrons and protons
respectively. $E'_{x,m}$ denotes the minimum injection energy of electrons or
 protons. $E'_{e,m}=m_e c^2 \bar\gamma'_p g(p) \frac{m_p}{m_e}
 \frac{\epsilon_e}{\epsilon_p}$ and $E'_{p,m}=\bar\gamma'_pm_pc^2g(p)$
 where $g(p)=\frac{p-2}{p-1}$ for $p>> 2$ and $g(p)\sim 1/6$ for
 $p=2$ \cite{gupta1}. $\bar\gamma'_p \sim 1$. The cooling break energy
 is denoted by $E'_{x,c}$. We have derived its expression by comparing
 the dynamical expansion time scale ($t'_{dyn}$) with the cooling time
 scale ($t'_{cool}$).
We use the general expression for $f_{\pi}(E_p)$ from \cite{nayan}
\begin{equation}
f_{\pi}(E_p) =f_{0} \left\{\begin{array}{l@{\quad \quad}l}
\frac{1.34^{\alpha_2-1}}{\alpha_2+1}(\frac{E_{p}}{E_{pb}})^{\alpha_2-1}
& E_p<E_{pb}\\\frac{1.34^{\alpha_1-1}}{\alpha_1+1}
(\frac{E_{p}}{E_{pb}})^{\alpha_1-1} & E_{p}>E_{pb}\end{array} \right.
\label{fpi}
\end{equation} 

Variability time, peak energy of the low energy (KeV-MeV) photon
fluence in MeV and corresponding threshold energy of protons for
$p\gamma$ interactions are denoted by $t_v$, $E_{\gamma,peak,MeV}$ and
$E_{pb}$ respectively. $\alpha_2=(p+2)/2$ is the spectral index of the
low energy photon spectrum above the peak energy $E_{\gamma,peak,MeV}$
and below it $\alpha_1=(p+1)/2$ (slow cooling).
\begin{eqnarray}
f_{0}^{HL}=\frac{0.729
L_{\gamma,51}}{\Gamma_{2.5}^4t_{v,-2}E_{\gamma,peak,MeV}}\frac{1}
{[\frac{1}{\alpha_2-2}-\frac{1}{\alpha_1-2}]}
\nonumber\\
f_{0}^{LL}=\frac{0.729 L_{\gamma,47}}{\Gamma_{1}^4t_{v,2}E_{\gamma,peak,keV}}
\frac{1}{[\frac{1}{\alpha_2-2}-\frac{1}{\alpha_1-2}]}
\label{f_0}
\end{eqnarray}
where $r=\Gamma^2 c t_v$ has been used. 
 The magnetic field in the comoving frame inside HL and LL GRBs can be expressed as \cite{zm02}
\begin{eqnarray}
B^{'HL} =5\times10^{4} {\rm G} \frac{(\xi_{1} \epsilon_{B,-1}
L_{tot,51})^{1/2}} {{\Gamma}_{2.5}^{3} t_{v,-2}}
\nonumber\\
B^{'LL}=1.5\times10^3 {\rm G} \frac{(\xi_{1}\epsilon_{B,-1}L_{tot,47})^{1/2}}{{\Gamma_1}^3 t_{v,2}}
\label{B}
\end{eqnarray}
where $\xi$ is the compression ratio, which is about 7 for strong
shocks. Comparing $t'_{dyn}=\frac{r}{\Gamma c}$
 and $t'_{syn}=\frac{6\pi m_p^2c^3}{\beta_p^{'2}\sigma_{p,T}E_pB^{'2}}$ the cooling break due to synchrotron emission can be derived
\begin{eqnarray}
E_{p,c,syn}^{HL,LL}=\frac{1}{t_v}\frac{3m_p^2c^4}{4\sigma_{p,T}{\beta'_p}^2cU\epsilon_B}
\nonumber\\
E_{p,c,syn}^{HL}=2\times10^{19}\frac{\Gamma_{2.5}^6 t_{v,-2}}{\xi_1\epsilon_{B,-1}L_{tot,51}}eV
\nonumber\\
E_{p,c,syn}^{LL}=2\times10^{18}\frac{\Gamma_1^6 t_{v,2}}{\xi_1 \epsilon_{B,-1}L_{tot,47}} eV
\label{p_cool}
\end{eqnarray}
where $\beta'_p\sim1$.\\
From $t'_{dyn}=t'_{p\gamma}$ one gets
\begin{eqnarray}
E_{p,c,p\gamma}^{HL,LL}=\frac{E_{pb}^{HL,LL}}{1.34}
\Big[\frac{(1+\alpha_1)}{f_0^{HL,LL}}\Big]^{1/(\alpha_1-1)}
\nonumber\\
E_{pb}^{HL}=3\times10^{16} \frac{\Gamma_{2.5}^2}{E_{\gamma,peak,MeV}}eV
\nonumber\\
E_{pb}^{LL}=3\times10^{16}\frac{\Gamma_1^2}{E_{\gamma,peak,keV}}eV
\end{eqnarray}
The luminosity emitted in synchrotron radiation by electrons is
\begin{equation}
L_{\gamma}=\frac{\eta_e \epsilon_e}{1+Y_e}L_{tot}
\end{equation}
$L_{tot}$ is the total luminosity of the GRB fireball. The radiation
efficiency of electrons
$\eta_e=[(E_{e,c}^{\prime}/E_{e,m}^{\prime})^{2-p},1]$ for slow, fast
cooling respectively.
The relative importance between inverse compton (IC) and synchrotron cooling of electrons is denoted by 
\begin{equation}
Y_e=\frac{L_{e,IC}}{L_{e,syn}}=\frac{-1+\sqrt{1+4\eta_e\epsilon_e/\epsilon_B}}{2}
\end{equation} 
where $L_{e,IC}$ and $L_{e,syn}$ (this is same as $L_{\gamma}$ in our discussion) are the luminosities of radiations by IC and 
synchrotron mechanisms respectively.
The proton flux in the source rest frame can be normalised as follows
\begin{equation}
\int_{E_{p,min}}^{E_{p,max}}E_p \frac{dN_p(E_p)}{dE_p} dE_p= \epsilon_p E_{tot}
\label{p_norm}
\end{equation}
$E_{p,min}=\Gamma E'_{p,m}$ is the minimum injection energy in the
source rest frame. $E_{tot}$ is the total energy of the fireball in
the same frame. The maximum energy of protons derived by comparing the acceleration and cooling (synchrotron and $p\gamma$ interaction) time scales are
$E'_{p,max,syn}$, $E'_{p,max,p\gamma}$ respectively.\\
$t'_{acc}=\frac{2\pi \zeta E'_p}{eB'c}=t'_{syn}$ and $\zeta\sim1$ give
\begin{eqnarray}
E_{p,max,syn}^{HL,LL}=\Gamma\Big[\frac{3m_p^2c^4e}{\sigma_{p,T}{\beta'_p}^2B'}\Big]^{0.5}\nonumber\\
E_{p,max,syn}^{HL}=10^{20}\Gamma_{2.5}^{5/2}\Big[\frac{t_{v,-2}}{(\xi_1 \epsilon_{B,-1}L_{tot,51})^{1/2}}\Big]^{1/2}eV\nonumber\\
E_{p,max,syn}^{LL}=6\times10^{19}\Gamma_1^{5/2}\Big[\frac{t_{v,2}}{(\xi_1 \epsilon_{B,-1}L_{tot,47})^{1/2}}\Big]^{1/2}eV\nonumber\\
\end{eqnarray}
and $t'_{acc}=t'_{p\gamma}$ gives
\begin{equation}
E_{p,max,p\gamma}=\Gamma\Big[\frac{1+\alpha_1}{f_0}\frac{eB'c\Gamma
t_v}{2\pi\zeta}
\Big(\frac{E'_{pb}}{1.34}\Big)^{\alpha_1-1}\Big]^{1/\alpha_1}
\end{equation}
For, $\alpha_1=1.5$
\begin{eqnarray}
E_{p,max,p\gamma}^{HL}=8\times10^{18}\Big[\frac{\xi_1\epsilon_{B,-1}L_{tot,51}}{E_{\gamma,peak,MeV}{f_0^{HL}}^2}\Big]^{1/3}eV
\nonumber\\
E_{p,max,p\gamma}^{LL}=3.6\times10^{18}\Big[\frac{\xi_1\epsilon_{B,-1} L_{tot,47}}{E_{\gamma,peak,keV}{f_0^{LL}}^2}\Big]^{1/3}eV
\end{eqnarray}
The maximum energy derived by comparing the acceleration and dynamical
 time scales is
\begin{eqnarray}
E_{p,max,dyn}^{HL,LL}=\frac{eB'c\Gamma^2 t_v}{2\pi}\nonumber\\
E_{p,max,dyn}^{HL}=6\times10^{19}\frac{(\xi_1\epsilon_{B,-1}L_{tot,51})^{1/2}}{{\Gamma_{2.5}}} eV
\nonumber\\
E_{p,max,dyn}^{LL}=2\times10^{19}\frac{(\xi_1 \epsilon_{B,-1} L_{tot,47})^{1/2}}{{\Gamma_1}} eV
\end{eqnarray}
One should also keep in mind that there are several limitations with the internal shock model. The thickness of the shell ($\frac{r}{\Gamma}$) may limit the maximum energy of the protons. Moreover, the relative kinetic energy between the colliding shells has to give the emission energy of GRBs. Hence, the radaition efficiency has to be high to attain the observed energy output during prompt emission of these sources. Also, the spectra observed by BATSE from some bright GRBs suggest at least within the same burst the peak photon energy distribution is narrow. This is hard to achieve within the internal shock model unless one invokes a strong bimodal distribution in the values of the bulk Lorentz factor. 
 \section{Proton Spectrum in the External Shock Model}
The expanding shells of GRBs push the external medium which results in external shocks. In \cite{rees2} the authors proposed that thermalization of
 the spectrum of the GRB can be avoided by introducing a slight
 contamination of baryons within the original fireball. The baryon
 loading parameter is defined as
\begin{equation}
\Gamma=\frac{E_{tot}}{Mc^2}
\end{equation}
where $M$ is the total baryon mass and $E_{tot}$ is the total energy
of the photon-pair fireball. \cite{mesz1} (MLR) showed that for a certain range of values of $\Gamma_0$ only a fraction of the initial energy is still in the form of radiation when the expanding fireball is optically thin. The rest
of the energy gets converted into kinetic energy of the baryons. For
high baryon loading the baryons can be accelerated such that almost
all the initial fireball energy transforms into kinetic energy of the
baryons.  After an initial period of acceleration, the fireball
expands freely. The energy in this phase is concentrated in a thin
slab initially of roughly constant thickness $\delta r \simeq r_0$.
After free expansion starts the shell thickness is given by
\cite{mesz1}
\begin{equation}
\delta r=\left\{ \begin{array}{l@{\quad \quad}l}
{{\Gamma}r_0} & r_s<r<r_b\\ r/{\Gamma} & r>r_b
\end{array}\right.
\end{equation}
where $r_s$ is the radius where free expansion starts and later the
shell starts expanding linearly with $r$ when $r>r_b$. The baryons can radiate if their kinetic energy can be properly randomized either by collisions with the external interstellar medium (ISM) or with slower/faster portions of the relativistic flow itself \cite{mesz2}. Finally, the fireball shell is decelerated by the ISM. During the initial interaction of the fireball shell with the
external medium a reverse shock propagates into the fireball to stop
it. A prompt fireball collects $1/\Gamma$ times its rest mass from the
ISM by the time the fireball's radius attends the deceleration radius.
The baryons or cosmic rays can loose energy by photopion interactions
$p+\gamma\rightarrow \pi^0,\pi^{+}$ and synchrotron radiation inside
the shell of thickness $\delta r$. 
We derive the cooling break energy and the maximum proton energy comparing the different time scale for the forward shock scenario. 
In the external shock model the expansion time scale at any time t is
$t'_{dyn}=\Gamma t$. At the deceleration epoch it is $\frac{r_{dec}}{c\Gamma_{dec}}$ where $r_{dec}$ is the deceleration radius and $\Gamma_{dec}$ is the Lorentz factor at that radius. This radius can be expressed as 
$r^{HL}_{dec}\simeq (2.5\times 10^{16}cm)(E_{tot,52}/n)^{1/3}(\Gamma_{dec,2.5})^{-2/3}$ \cite{zm04} and $r^{LL}_{dec}\simeq (5.4\times10^{16}cm) (E_{tot,50}/n)^{1/3}(\Gamma_{dec,1})^{-2/3}$ where
 $E_{tot}$ is the total energy of a blast wave. The comoving magnetic field at the deceleration epoch is $B^{'HL}_{dec}=c\Gamma_{dec}(32\pi n m_p \epsilon_B)^{1/2}=39 \Gamma_{dec,2.5}(n\epsilon_{B,-1})^{1/2}$G, $B^{'LL}_{dec}=1.23 \Gamma_{dec,1}(n\epsilon_{B,-1})^{1/2}$G. $n\sim 1$ is particle number density of the ambient medium.
 
For interstellar medium (ISM) the radius of wind expansion at any time $t$ measured in the source rest frame is (\cite{zm01})
\begin{eqnarray}
r^{HL}(t)=4\times10^{16}\Big(E_{tot,52}/n\Big)^{1/4} {t_h}^{1/4} cm 
\nonumber\\
r^{LL}(t)=1.25\times10^{16}\Big(E_{tot,50}/n\Big)^{1/4} {t_h}^{1/4} cm
\end{eqnarray}
The Lorentz factor decreases with time as follows 
\begin{eqnarray}
\Gamma^{HL}(t)=19.4 \Big(E_{tot,52}/n\Big)^{1/8}{t_h}^{-3/8}
\nonumber\\
\Gamma^{LL}(t)=11\Big(E_{tot,50}/n\Big)^{1/8}t_h^{-3/8}
\end{eqnarray}
where $t_h$ is the time measured in the source rest frame in hours.
We have used $r=\Gamma^2 c t$.
The time dependent comoving magnetic field is
\begin{eqnarray}
B^{'HL}(t)=7.5G \epsilon^{1/2}_B {E_{tot,52}}^{1/8} n^{3/8} {t_h}^{-3/8}
\nonumber\\
B^{'LL}(t)=4.2G\epsilon_B^{1/2} {E_{tot,50}}^{1/8}n^{3/8}t_h^{-3/8}
\end{eqnarray}
 From $t'_{dyn}=t'_{syn}$, the break energy in the proton spectrum of GRBs due to synchrotron cooling of protons
 is
\begin{eqnarray}
E^{HL}_{p,c,syn}=2\times10^{22} {E_{tot,52}}^{-1/4}n^{-3/4}{\epsilon_B}^{-1}t^{-1/4}_h eV\nonumber\\
E^{LL}_{p,c,syn}=6\times10^{22} {E_{tot,50}}^{-1/4}n^{-3/4}{\epsilon_B}^{-1}
t^{-1/4}_h eV
\end{eqnarray}
At the deceleration epoch this expression becomes 
\begin{eqnarray}
E^{HL}_{p,c,syn}=7.5\times10^{21}\frac{\Gamma_{dec,2.5}^{2/3}}{n\epsilon_{B,-1}}\Big(\frac{E_{tot,52}}{n} \Big)^{-1/3}eV\nonumber\\
\nonumber\\
E^{LL}_{p,c,syn}=1.64\times10^{21}\frac{\Gamma_{dec,1}^{2/3}}{n\epsilon_{B,-1}}\Big(\frac{E_{tot,50}}{n}\Big)^{-1/3}eV\nonumber\\
\end{eqnarray}
 $t'_{dyn}=t'_{p\gamma}$ gives
\begin{equation}
E^{HL,LL}_{p,c,p\gamma}=\frac{E^{HL,LL}_{pb}}{1.34}\Big[\frac{(1+\alpha_1)}{f^{HL,LL}_0}\Big]^{\frac{1}{(\alpha_1-1)}}
\end{equation}

 The time dependence of this cooling break energy is calculated below.
 The photon energy flux at the peak/break energy is proportional to  
$t^{-1},t^{-p/2}$ in the case of fast and slow  cooling respectively. For $p\sim 2$ the two cases are almost similar. This is also true for the luminosity of the low energy photons if the threshold energy of photo-pion production is below the peak energy. If the threshold energy is above the peak energy then the luminosity would be proportional to  $t^{-p/4-1/2}$ for both fast and slow cooling of electrons.
The threshold energy condition for $p\gamma$ interaction in the source rest frame gives $E_{\gamma,peak}(t)E_{pb}(t)=0.3\Gamma^2(t)GeV^2$. 
If it is slow cooling $E_{\gamma,peak}(t)\propto t^{-1/2}$ and in the case of fast cooling it is proportional to $t^{-3/2}$. If the threshold energy of $p\gamma$ interaction is lower than the peak fluence energy in the afterglow photon spectrum then luminosity $L(t)\propto t^{-1},t^{-p/2}$ in the case of fast and slow cooling respectively. For $p\sim2$ we get $L(t)\propto t^{-1}$ in both the cases. Considering the external medium to be ISM $f^{HL,LL}_0\propto L(t)/( \Gamma^{4}(t) t E_{\gamma,peak}(t))\propto t^{0},t$ for slow and fast cooling respectively. The break energy due to $p\gamma$ cooling in the source rest frame  
 is proportional to $t^{-1/4},t^{-5/4}$ with $\alpha_1\approx1.5$, for slow and fast cooling respectively. The time dependence of the $p\gamma$ cooling depends on several factors like the photon flux, the external medium (ISM or wind) and whether electrons cool slow or fast. For different underlying assumptions one gets different time dependence  of the $p\gamma$ cooling (\cite{wang,murase}).
The cooling break energy in the proton spectrum is $E'_{p,c}=min[E'_{p,c,syn},E'_{p,c,p\gamma}]$.
We compare the cooling and acceleration time scales to derive the maximum energy of protons.
\begin{eqnarray}
E^{HL}_{p,max,syn}=7\times10^{21} \frac{\Gamma_{dec,2.5}^{1/2}}{(n \epsilon_{B,-1})^{1/4}}eV
\nonumber\\
E^{LL}_{p,max,syn}=4\times10^{20}\frac{\Gamma_{dec,1}^{1/2}}{(n \epsilon_{B,-1})^{1/4}}eV
\end{eqnarray}
After the deceleration epoch the time dependent maximum energy obtained by comparing synchrotron cooling and acceleration time scales is
\begin{eqnarray}
E^{HL}_{p,max,syn}=\Gamma(t)\Big[\frac{3e m_p^2 c^4}{\sigma_{p,T} B^{'HL}(t)}\Big]^{0.5}=6\times10^{20} \epsilon_B^{-1/4} E_{tot,52}^{1/16} n^{-5/16} t_h^{-3/16} eV 
\end{eqnarray}
for $\beta'_p \sim 1$.
The maximum energy of protons derived by comparing the acceleration and  pion production time scales is
\begin{equation}
E^{HL}_{p,max,p\gamma}=\Gamma(t)\Big[\frac{eB^{'HL}(t)}{2\pi }\frac{\Gamma(t) t (1+\alpha_1)}{f^{HL}_0}\Big(\frac{E^{'HL}_{pb}}{1.34}\Big)^{\alpha_1-1}\Big]^{1/\alpha_1}
\end{equation}
At the deceleration epoch 
\begin{equation}
E^{HL}_{p,max,p\gamma}=\Gamma_{dec}\Big[\frac{eB^{'HL}_{dec}}{2\pi }\frac{r^{HL}_{dec}(1+\alpha_1)}{\Gamma_{dec}f^{HL}_0}\Big(\frac{E^{'HL}_{pb}}{1.34}\Big)^{\alpha_1-1}\Big]^{1/\alpha_1}
\end{equation}
for $\alpha_1=1.5$
\begin{eqnarray}
E^{HL}_{p,max,p\gamma}=5\times 10^{18} \Big[\frac{n \epsilon_{B,-1}}{(f^{HL}_0)^2} \Big(\frac{E_{tot,52}}{n}\Big)^{2/3} \frac{(\Gamma_{dec,2.5})^{8/3}}{E_{\gamma,peak,MeV}}\Big]^{1/3} eV
\nonumber\\
E^{LL}_{p,max,p\gamma}=10^{18}\Big[\frac{n \epsilon_{B,-1}}{(f^{LL}_0)^2} \Big(\frac{E_{tot,50}}{n}\Big)^{2/3} \frac{(\Gamma_{dec,1})^{8/3}}{E_{\gamma,peak,keV}}\Big]^{1/3}eV
\end{eqnarray}
From $t'_{dyn}=t'_{acc}$ we get
\begin{eqnarray}
E^{HL,LL}_{p,max,dyn}=\frac{eB^{'HL,LL}(t)c}{2\pi} \Gamma(t)^{2} t\nonumber \\
E^{HL}_{p,max,dyn}=1.4\times10^{19} {\epsilon_B}^{1/2}(E_{tot,52}^3 n/ t_h)^{1/8} eV\nonumber\\
E^{LL}_{p,max,dyn}=2.5\times10^{18} {\epsilon_B}^{1/2}(E_{tot,50}^3 n/ t_h)^{1/8} eV
\end{eqnarray}
At the deceleration epoch
\begin{eqnarray}
E^{HL,LL}_{p,max,dyn}=\frac{e B^{'HL}_{dec}}{2 \pi }r^{HL}_{dec}
\nonumber\\
E^{HL}_{p,max,dyn}=2\times 10^{19} (n \epsilon_{B,-1})^{1/2}\Big(\frac{E_{tot,52}}{n}\Big)^{1/3} (\Gamma_{dec,2.5})^{1/3} eV\nonumber\\
E^{LL}_{p,max,dyn}=1.5\times10^{18} (n\epsilon_{B,-1})^{1/2}\Big(\frac{E_{tot,50}}{n}\Big)^{1/3}(\Gamma_{dec,1})^{1/3}eV
\end{eqnarray}

We calculate $E_{p,max}$ as discussed in the previous section.
The protons lose energy as long as the cooling time scale is much shorter than the wind expansion time scale. 
For $\epsilon_p\sim1$ the mean Lorentz factor of the protons is roughly the Lorentz factor of the forward shock. 
The total number of protons increases as
 $nr(t)^3\propto t^{3/4}$ incase of ISM and as $n r(t)^3\propto r(t)\propto t^{1/2}$ for wind medium. The total energy carried by protons remains constant, 
$\epsilon_pE_{tot}=N_p \Gamma(t)^2 m_p c^2\propto t^0$. The proton spectrum has been normalised as given in eqn.(\ref{p_norm}). It is important to discuss the 
 difficulty with the external shock model. The particles (e,p) need to be thrown back to the external shock region by some magnetic field for their reacceleration. The magnetic field in the ISM is known to be low. For acceleration of the  protons to very high energy in the external shock model the ambient magnetic field has to be more than its typical value in the ISM. It is not physically impossible that the ambient medium close to GRBs is more magnetized than the average  ISM. Alternatively, magnetic field amplification may lead to acceleration of particles to very high energy. For discussions on particle acceleration in this way
 in the non relativistic regime one may see \cite{blasi1,blasi2}.       
 
 \section{Internal Shocks:Average Proton Flux from GRBs}
We have used the broken power law luminosity distribution functions
for HL GRBs \cite{liang}.
\begin{equation}
S(L)=S_0\Big[\Big(\frac{L}{L_b}\Big)^{\beta_1}+\Big(\frac{L}{L_b}\Big)^{\beta_2}\Big]^{-1}
\end{equation}
The break luminosity $L_b=10^{52}$ erg/sec for HL GRBs. The values of $\beta_1$, $\beta_2$ are taken to be 0.65, 2.3. The star formation rate is
(\cite{por})
\begin{equation}
R_{GRB}=23\rho_0\frac{e^{3.4z}}{e^{3.4z}+22.0}
\end{equation}
and $\rho_0\sim 1$. The diffuse proton flux
can be calculated as follows
\begin{equation}
M^{ob}_{p}(L_{\gamma},\Gamma,T_d,t_v,E^{ob}_p)=\int_0^{z_{max}}
\frac{dN^{ob}_{p}(E^{ob}_{p})}{dE^{ob}_{p}}\frac{R_{GRB}(z)}{1+z}dV(z)
\end{equation}
$T_d$ is the duration of a burst and $t_v$ is the variability time scale in the internal shock model.
The isotropic energy emitted
by low energy photons $(E_{\gamma}=T_{d}L_{\gamma})$ is related to
the total energy of a GRB
\begin{equation}
E_{\gamma}=\frac{\epsilon_e \eta_e}{1+Y_e} E_{tot}
\end{equation}
Amati-relation \cite{amati,ghir} has been used to calculate the peak fluence energy in the low energy photon spectrum while using the
internal shock model. We have averaged over all the relevant parameters ($\Gamma$, $L_{\gamma}$, $T_d$, $t_v$) as discussed in \cite{nayan}. For the log normal distribution in Lorentz factor the peak is assumed to be at $log(\Gamma_m)=2.6$ with standard deviation 0.3 for HL GRBs. The durations of the bursts are assumed to follow log normal distribution with the mean at $log(T_{d,m})=1.5$ with standard deviation 0.5. The mean of the log normal distribution in variability time is assumed to be at $log(t_{v,m})=-1.5$ with standard deviation 0.3. In Fig.1. our calculated proton spectrum has been compared with PA data above 60EeV. If we increase the average deflection angle of the cosmic ray protons from $4^{\circ}$ to $10^{\circ}$ the proton flux does not change significantly.
\begin{figure*}[t]
\begin{center}
\centerline{
\epsfxsize=10.cm\epsfysize=7.cm
\epsfbox{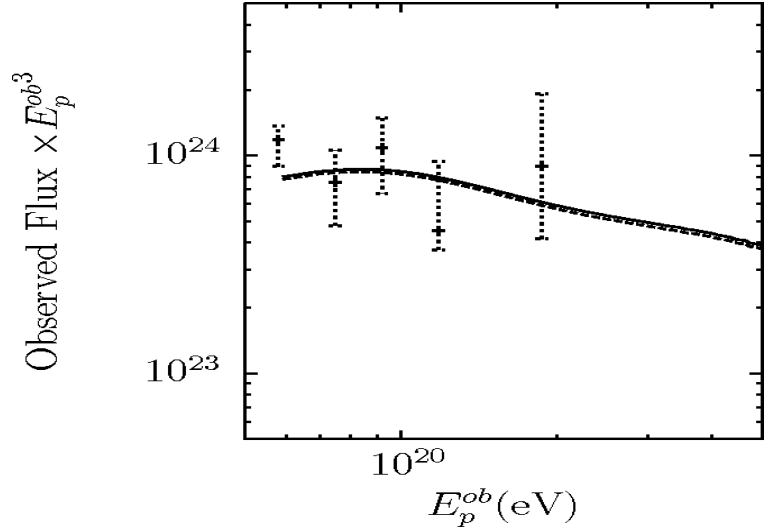} }
\end{center}
\caption{proton flux above 60EeV in $eV^2 m^{-2} sec^{-1} sr^{-1}$ from HL GRBs using the internal shock model, $z_{max}=0.02$, $\theta_a=4^{\circ},10^{\circ}$(solid line, dashed line).We have considered fast cooling of electrons and $p=2.3$ \cite{gupta1}, $\epsilon_p=0.8$ and $\epsilon_e=0.1$. The data points are from recent results of PA experiment \cite{rec}}
\end{figure*}
\section{External Shocks:Proton Flux from GRBs}
In the external shock model we have calculated the proton flux from all GRBs using log normal distribution in total energy with the mean at $log(E_{tot,m,erg})=54$ and standard deviation 2 \cite{liang1}. The distribution in the values of the Lorentz factor at the deceleration epoch is also assumed to be log normal with the mean 2.6 and standard deviation 0.3. The proton flux from all HL GRBs at the deceleration epoch is shown in Fig.2. We have used the particle number density in the ISM to be $n=1cm^{-3}$. After the deceleration epoch the maximum energy of protons decreases with time and the chances of getting cosmic ray protons above 60 EeV becomes less. From LL GRBs we expect even lower cosmic ray flux than that from HL ones.  
\begin{figure*}[t]
\begin{center}
\centerline{
\epsfxsize=10.cm\epsfysize=7.cm
\epsfbox{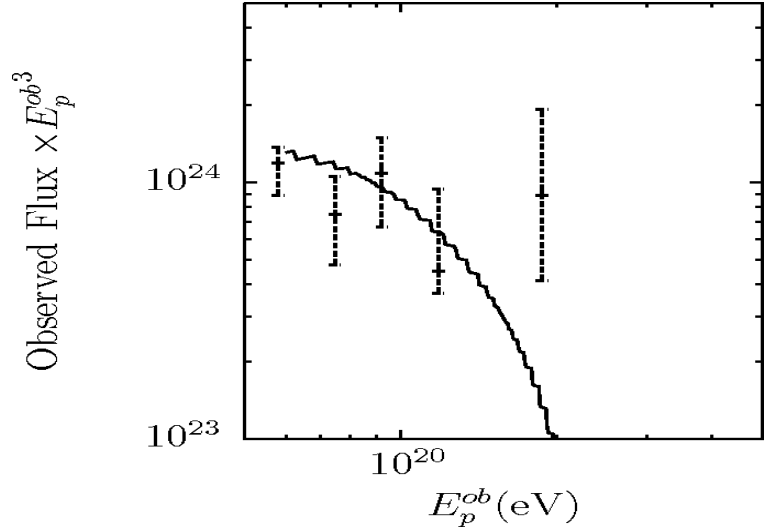} }
\end{center}
\caption{proton flux above 60EeV in $eV^2 m^{-2} sec^{-1} sr^{-1}$ from HL GRBs using the external shock model at the deceleration epoch, $z_{max}=0.02$, $\theta_a=4^{\circ}$ as before also $p=2.3$ \cite{gupta1}. $\epsilon_p=0.5$ and $\epsilon_B=0.3$.}
\end{figure*}
\section{Maximum Energies of Protons:Internal vs External Shocks}  
In Fig.3. the three plots show how the maximum and cooling energies depend on the GRB parameters in the internal shock model. For $\Gamma\sim100$ the minimum of the maximum energies of protons is due to $p\gamma$ cooling. As Lorentz factor increases the maximum energy of the protons is determined by the dynamical time scale. If the total energy of the low energy photons is varied synchotron cooling of protons remains unaffected as long as the luminosity of a GRB remains constant. For $\Gamma=300$, $t_v=1sec$ and $L_{\gamma}=10^{51}eg/sec$ the dynamical timescale determines the maximum energy of the protons in the entire range of
 $E_{\gamma}$ considered in our work. The maximum energy determined by the dynamical time scale is independent of the variability time of GRBs as shown in the third plot of Fig.3. For higher values of variability times of GRBs the rate of dynamical expansion determines the minimum of the maximum energies of protons. 
Fig.4. shows the various energies (maximum and cooling) in the case of external shocks. Unless Lorentz factor is nearly 1000, $p\gamma$ cooling determines the 
 maximum energies of protons. $p\gamma$ cooling is more important in the case of external shocks than in internal shocks. We vary the total energy of the GRB, particle density of the medium and find $p\gamma$ cooling gives the minimum of the maximum energies of potons. The cooling break energies are not important in case of external shocks as they are very high.    
\begin{figure*}[t]
\begin{center}
\centerline{
\epsfxsize=6.cm\epsfysize=6.cm
\epsfbox{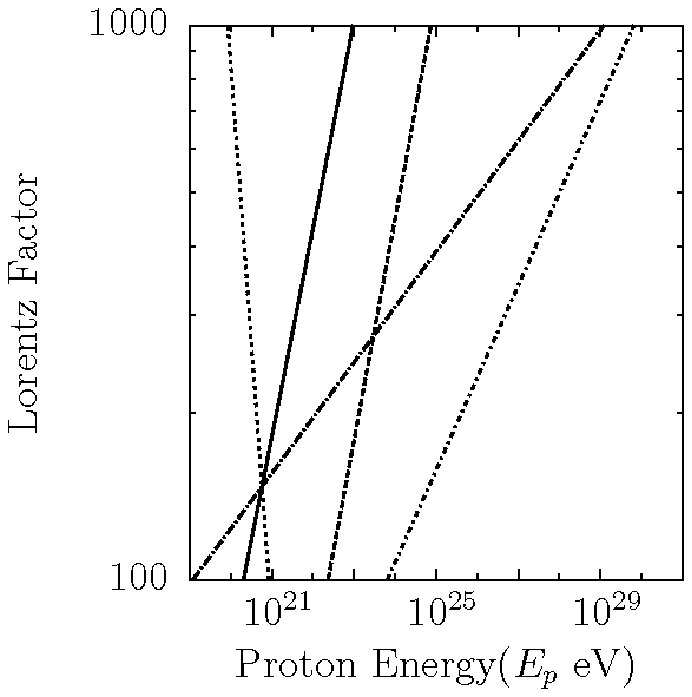}
\epsfxsize=6.cm\epsfysize=6.cm
\epsfbox{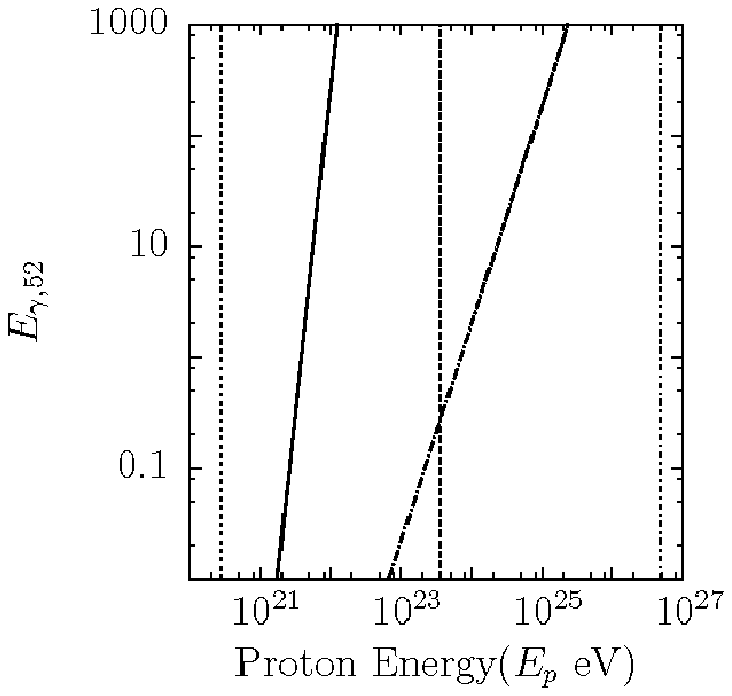}
\epsfxsize=6.cm\epsfysize=6.cm
\epsfbox{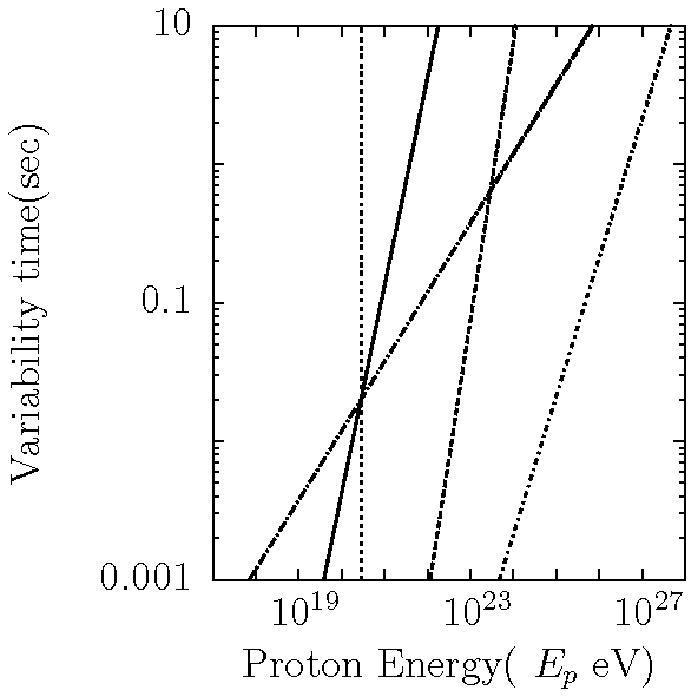}}
\end{center}
\caption{Internal Shock Model: First plot- total energy carried by low energy photons $E_{\gamma,52}=1$, $t_v=1sec$; second plot-$\Gamma=300$, $t_v=1sec$; third plot- $\Gamma=300$, $E_{\gamma,52}=1$. $E_{p,max,p\gamma}$ (solid line); $E_{p,max,dyn}$ (dotted line); $E_{p,max,syn}$(dashed line); $E_{p,c,p\gamma}$ (long dash dotted line); $E_{p,c,syn}$ (small dash dotted line); $\epsilon_p=0.8, \epsilon_e=0.1$, $L_{\gamma}=10^{51}erg/sec$}
\end{figure*}

\begin{figure*}[t]
\begin{center}
\centerline{
\epsfxsize=6.cm\epsfysize=6.cm
\epsfbox{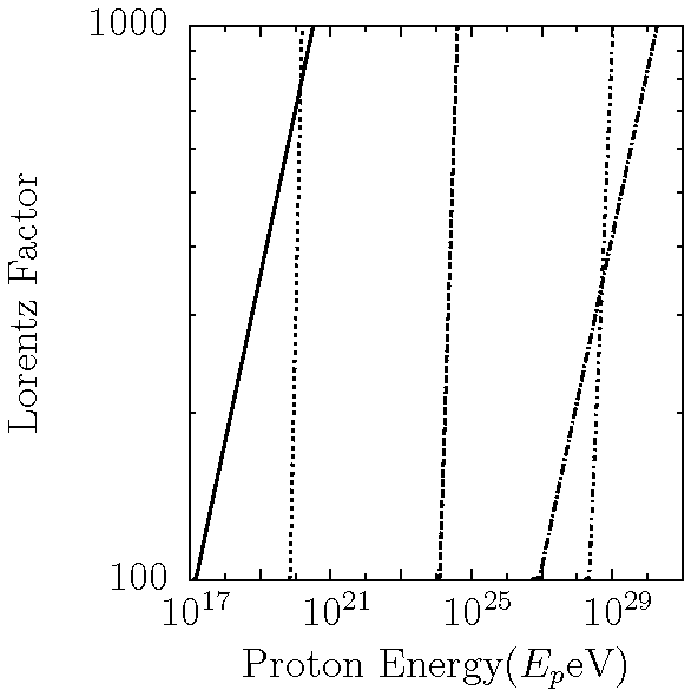}
\epsfxsize=6.cm\epsfysize=6.cm
\epsfbox{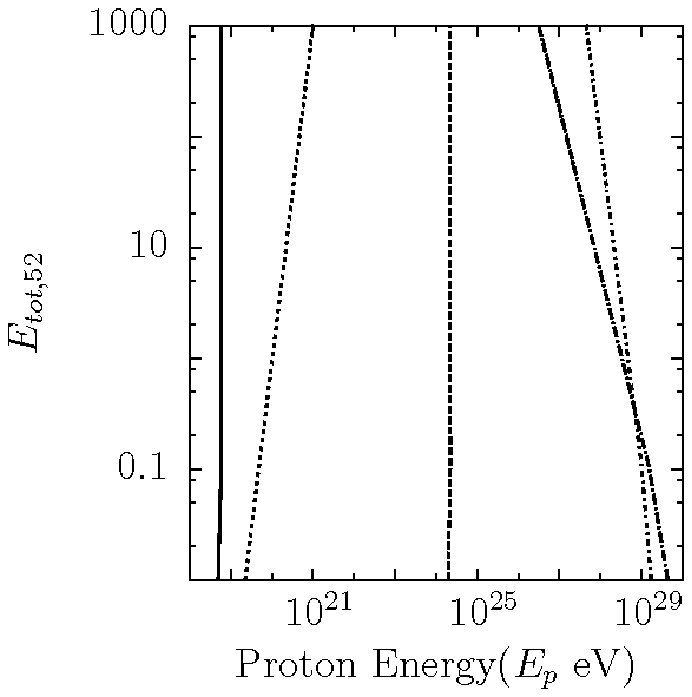}
\epsfxsize=6.cm\epsfysize=6.cm
\epsfbox{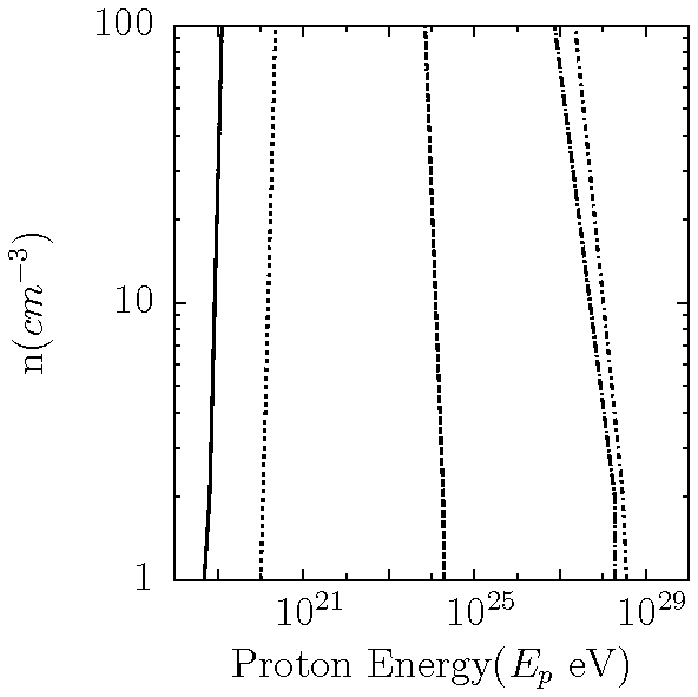}}
\end{center}
\caption{External Shock Model: First plot- $E_{tot,52}=1$, $n=1cm^{-3}$; second plot- $\Gamma=300$, $n=1cm^{-3}$; third plot- $\Gamma=300$, $E_{tot,52}=1$; $\epsilon_B=0.5, \epsilon_p=0.3$. Different line styles used as in Fig.3.}
\end{figure*}
\section{Conclusions}
The leading theoretical models often used to explain emissions from GRBs are the internal and the external shock models. Within these models cosmic rays can be accelerated to extremely high energies inside these sources. Cosmic ray proton fluxes above 60 EeV have been calculated from GRBs distributed upto a redshift of 0.02 assuming distributions in the values of the relevant GRB parameters. The recent data from the Pierre Auger experiment is compared with our calculated proton fluxes. Our results show that GRBs can emit cosmic ray flux comparable to the PA data above 60EeV both in the internal and external shock models. In the internal shock model $p\gamma$ cooling and the dynamical expansion time scale of the GRBs are important in determining the maximum energy of the protons while in the case of the external shock model $p\gamma$ cooling determines the maximum energy of the shock accelerated protons.   
\section{Acknowledgement}
I thank Bing Zhang for many helpful communications and to the anonymous referee for important suggestions. 

\end{document}